\documentclass[twocolumn,amsfonts,amsmath,amssymb,prb]{revtex4}
\usepackage{graphicx}
\usepackage{bm}

\begin{document}

\title{Determinations of upper critical field in continuous Ginzburg-Landau model} 

\author{L. Wang}
\author{H. S. Lim}
\author{C. K. Ong}

\address{Center for Superconducting and Magnetic Materials and 
Department of Physics, 
National University of Singapore, 
2 Science Drive 3, 
Singapore 117542. 
} 

\begin{abstract}
Novel procedures to determine the upper critical field $B_{c2}$ 
have been proposed within a continuous Ginzburg-Landau model. 
Unlike conventional methods, where $B_{c2}$ is obtained through 
the determination of the smallest eigenvalue of an appropriate eigen equation, 
the square of the magnetic field is treated as eigenvalue problems 
so that the upper critical field can be directly deduced. 
The calculated $B_{c2}$ from the two procedures are consistent with each other 
and in reasonably good agreement with existing theories and experiments. 
The profile of the order parameter associated with $B_{c2}$ is 
found to be Gaussian-like, further validating the methodology proposed. 
The convergences of the two procedures are also studied. 
\end{abstract}

\maketitle

\section{introduction}

The determinations of the upper critical field $B_{c2}$ are 
based on the Ginzburg-Landau-Abrikosov-Gorkov\cite{Ginzburg50,Abrikosov57,Gorkov59} 
(GLAG) framework that is applicable to practically all superconductors. 
The starting point of the macroscopic description is 
the first Ginzburg-Landau (GL) equation\cite{Ginzburg50,Abrikosov57} 
or its similarities. Examples  would be  
the standard linearized GL equation\cite{Cheng99}, 
the harmonic oscillator equation\cite{Abrikosov57,Tinkham96,Duzer99,Jin89},
the Mathieu equation\cite{Klemm75,Bulaevskii76,Schneider93}, 
the continuous Ginzburg-Landau equation\cite{Koyama92}, 
the GLd equation(s)\cite{Joynt90,Berlinsky95,Franz96,Chang98} 
[Ginzburg-Landau model for (d+s)-wave superconductors], and 
the Ginzburg-Pitaevskii equation\cite{Alexandrov96}. 

On the other hand, the linear gap equation\cite{Gorkov59,Gennes64,Werthamer63} 
is adopted to  microscopically determine the upper critical field. 
This equation can be converted into an equation including the digamma function. 
The parameter of the digamma function is related to the smallest eigenvalue of 
an eigen equation with appropriate boundary conditions. Having obtained 
the smallest eigenvalue of the eigen equation, $B_{c2}$ may then be 
implicitly determined from the equation including 
the digamma function ($B_{c2}$ is in connection with 
the smallest eigenvalue). Such a procedure or its extensions may be seen, 
for example, in the calculations of $B_{c2}$ for 
type-II superconductors\cite{Maki64,Werthamer66} (s-wave), 
heavy-fermion superconductors\cite{Scharnberg80,Lukyanchuk87} (p-wave), 
high $T_c$ superconductors\cite{Rieck89,Maki97} [(d+s)-wave], 
layered superconductors\cite{Klemm75,Ovchinnikov96,Lebed98}, 
superlattices\cite{Takahashi86,Yuan91}, 
thin films\cite{Yuan94,Smith00}, multilayers\cite{Kuwasawa96} and 
organic superconductors\cite{Huang89}.

It is tempting to conclude that almost all the efforts 
\cite{Abrikosov57,Maki64,Werthamer66,Scharnberg80,
Lukyanchuk87,Rieck89,Maki97,Klemm75,Ovchinnikov96,Lebed98,Takahashi86,
Yuan91,Yuan94,Smith00,Kuwasawa96,Huang89,Cheng99,Tinkham96,Duzer99,Jin89,
Bulaevskii76,Schneider93,Koyama92,Joynt90,Berlinsky95,Franz96,Chang98} 
made to obtain $B_{c2}$ are via the determination of the smallest eigenvalue of 
an appropriate eigen equation. For instance, in Ref.~\onlinecite{Duzer99}, 
$B_{c2}$ is inversely proportional to the smallest eigenvalue 
(the lowest Landau level) of the harmonic oscillator equation. 

Nevertheless, for all the GL-like equations as well as the linear gap equation, 
one may have a universal definition for the upper critical field: 
it is the maximum magnetic field at which the corresponding equation has 
a nontrivial solution\cite{exception}. Such a nontrivial solution would 
naturally correspond to the eigen function associated with 
the smallest eigenvalue and would correspond to the solution to 
the linear gap equation at $B_{c2}$\cite{reenter}. 
It is worth noting that the eigen function associated with 
the smallest eigenvalue may be just one of the solutions to 
the linear gap equation (see, for example, 
Refs.~\onlinecite{Gennes64}, \onlinecite{Werthamer66} and \onlinecite{Gorkov76}). 

Several approaches, for example, the perturbation method\cite{Jin89,Joynt90,Chang98} 
and the variational technique\cite{Berlinsky95,Franz96}, have been adopted to 
find $B_{c2}$ by obtaining the smallest eigenvalue. We note that 
in Ref.~\onlinecite{Cheng99}, the authors iteratively calculated 
the upper critical field of a thin-film superconductor with a ferromagnetic dot: 
first guessing a value for the upper critical field, then solving the standard 
linearized GL equation to make sure the calculated value of the upper critical field 
is equal to the guessed one. Otherwise, equate the newly calculated value to 
the guessed one and resolve the linear equation until the two values are equal to 
each other. It can be seen that such an iterative method is still related to 
obtaining the smallest eigenvalue\cite{Cheng99}.

In this paper, instead of obtaining the smallest eigenvalue, from which $B_{c2}$ 
is implicitly or indirectly determined, we will directly calculate $B_{c2}$: 
the square of the magnetic field $B^2$ is treated as an eigenvalue of 
an eigen equation so that $B_{c2}$ can be directly deduced from 
the largest eigenvalue of the eigen equation. Within a continuous 
Ginzburg-Landau model\cite{Wang01a}, two procedures will be presented to 
obtain the corresponding eigen equations, either of which can determine $B_{c2}$. 
Note that the applications of these procedures have been published in 
Refs.~\onlinecite{Wang01b} and \onlinecite{Wang02a}.

\section{Model}

The continuous Ginzburg-Landau model (CGL) is applicable to 
layered superconductors\cite{Wang01a}. The unit cell describing 
the layering structure consists of alternating superconducting (S) and 
insulating (I) layers. The $z$ axis is normal to the layers and 
its origin is at the midpoint of one of the I layers. The center of 
the S layer is located at $D/2$, where $D$ is the size of 
the unit cell\cite{Wang02a}. The GL coefficients and 
the effective superpair masses (perpendicular and parallel to the layers) 
in the CGL free energy are spatially dependent. The CGL free energy is,  
\begin{widetext}
\begin{eqnarray}
F & = & \int d\vec{r} \int dz \left[ \alpha(T,z) |\Psi(\vec{r},z)|^{2} +
\frac{1}{2} \beta |\Psi(\vec{r},z)|^{4} + \frac{\hbar^{2}}{2M(z)} 
\left| \left( \frac{\partial}{\partial z} - 
\frac{2ie}{\hbar}A_{z} (\vec{r},z)\right) 
\Psi(\vec{r},z) \right|^{2} \right. \nonumber \\
& & \mbox{} + \left. \frac{\hbar^{2}}{2m(z)}\left| \left( \nabla^{(2)} -  
\frac{2ie}{\hbar}\vec{A}^{(2)}(\vec{r},z)\right)\Psi(\vec{r},z)\right|^{2}
+ \frac{1}{2\mu_{0}}B^{2}(\vec{r},z) \right], 
\label{eq:freeE}
\end{eqnarray}
\end{widetext}
where $T$ is the temperature, $\vec{r}=(x, y)$ is the planar vector and 
$\vec{A}(\vec{r},z) = (\vec{A}^{(2)}(\vec{r},z), A_{z}(\vec{r},z))$ is 
the vector potential. The effective masses and the GL condensation coefficient 
$\alpha(T,z)$ are assumed to be periodic with a period $D$. $\beta$, however, 
is held fixed as it does not affect the qualitative behavior of 
the studied system\cite{Kleiner97}. As before\cite{Wang01b,Wang01a,Wang02a,Wang02b}, 
we choose 
\begin{subequations} \label{eq:coeficient}
\begin{eqnarray}
\alpha (T,z)& =&  \left[ \alpha_0 + \alpha_{1}\cos (2\pi z/D)\right] (1-T/T_{c}),  
\label{eq:coeficienta} \\ 
\frac{1}{M(z)} & = & G_{0} + G_{1} \cos (2\pi z/D), 
\label{eq:coeficientb} \\
\frac{1}{m(z)} & = &  g_{0} + g_{1} \cos (2\pi z/D), 
\label{eq:coeficientc}
\end{eqnarray}  
\end{subequations}
where $T_c$ is the transition temperature. $\alpha_0$, $\alpha_1$, $G_0$, $G_1$, 
$g_0$ and $g_1$ are model parameters. Note that for dirty materials, 
the linear GL theory for $B_{c2}$ may be extended to 
low temperatures\cite{Maki64,Tinkham96,Gorkov76}.

Let an external field $B$ be applied along the $y$-direction so that 
the vector potential can be taken as $\vec{A}=(Bz, 0, 0)$. Assuming 
$\Psi(\vec{r},z)= e^{i\vec{k_{\parallel}} \cdot \vec{r}}\Psi(z)$ and 
minimizing the free energy of Eq.~\ref{eq:freeE}, we get
\begin{eqnarray}
&&-\frac{\hbar^{2}}{2M(z)}\frac{\partial^{2}}{\partial z^{2}}\Psi(z) -
\frac{\hbar^{2}}{2}\left[ \frac{\partial}{\partial z} \frac{1}{M(z)}
\right]\frac{\partial}{\partial z} \Psi(z) + \nonumber \\
&&\left[ \frac{1}{2m(z)}(2eB)^{2}
(z-z_s)^{2} + \frac{\hbar^{2}k_{y}^{2}}{2m(z)}\right] \Psi(z) + \nonumber \\
&&\alpha(T,z)\Psi(z)+\beta|\Psi(z)|^{2}\Psi(z)  =  0,
\label{eq:1D}
\end{eqnarray}
with $z_s= \hbar k_{x}/(2eB)$. At $B=B_{c2}$, the order parameter is 
small enough so that the term $\beta|\Psi(z)|^{2}\Psi(z)$ in Eq.~\ref{eq:1D} can 
be omitted. The superconducting order at $B_{c2}$ nucleates in the S layer first 
so that one may choose $z_s=D/2$\cite{Koyama92}. To explore the features of 
the order parameter along the $z$-direction, one may assume $k_y = 0$. Finally, 
we arrive at the following equation,
\begin{eqnarray}
&&-\frac{\hbar^{2}}{2M(z)}\frac{\partial^{2}}{\partial z^{2}}\Psi(z) -
\frac{\hbar^{2}}{2}\left[ \frac{\partial}{\partial z} \frac{1}{M(z)}
\right] \frac{\partial}{\partial z} \Psi(z) + \nonumber \\
&&\left[ \alpha(T,z)+\frac{1}{2m(z)}(2eB)^{2}(z-\frac{D}{2})^{2} \right] \Psi(z) 
= 0. 
\label{eq:eq1DFinal}
\end{eqnarray}
At a given temperature $T$, the maximum magnetic field $B$ at which 
a nontrivial solution satisfies the above equation gives a point in 
the $B_{c2}$-$T$ plot. Eq.~\ref{eq:eq1DFinal} will be numerically solved 
subject to the following boundary conditions, 
\begin{subequations} \label{cond}
\begin{eqnarray}
&& \Psi(0) = \Psi(D),  \\ 
\label{conda} 
&& \hspace{-3mm} \left. \frac{\partial}{\partial z} \Psi(z)\right|_{z=D} =  0. 
\label{condb}
\end{eqnarray}  
\end{subequations}
Note that neglecting the size effect in the $z$-direction and considering 
the layering structure and the superposition of different superpairs as 
a mean field effect, the variation of the macroscopic 
wave function-the order parameter along the $z$-direction may 
have a periodic property (for example, see Ref.~\onlinecite{Koyama92}). 
It should be mentioned that the new methodology of determining 
the upper critical field, which is to be described below, 
is also applicable to other boundary conditions\cite{Wang02a,Wang02b}.

\section{Methodology}

\subsection{Procedure I}

Taking into account the boundary conditions of Eq.~\ref{cond}, 
Eq.~\ref{eq:eq1DFinal} can be transformed into a system of 
equations which can be simplified to 
\begin{equation}
\textbf{U}\bm{\Psi}=0, 
\label{Ueq}
\end{equation}
where $\bm{\Psi}=(\Psi_1, \Psi_2, ..., \Psi_{2n+1})^\prime$ is 
a column vector representing the discrete solutions of 
Eq.~\ref{eq:eq1DFinal}. The symbol $^\prime$ here indicates transpose and 
$n$ is a positive integer. $\textbf{U}$ is a $(2n+1)\times(2n+1)$ 
sparse matrix  having the following structure\cite{Wang02b}, 
\begin{widetext}
\begin{eqnarray}
{\scriptsize  
\left(
\begin{array}{l} \vspace{1.5mm} 
\hspace{1mm} 1  \hspace{138mm} -1 \\ \vspace{1.5mm}
B_{2}\hspace{3mm} C_{2}\hspace{3mm} D_{2}\hspace{3mm} E_{2} \hspace{109mm} A_{2} 
\\ \vspace{1.5mm}
A_{3}\hspace{3mm} B_{3}\hspace{3mm} C_{3}\hspace{3mm} D_{3}\hspace{3mm} E_{3}   
\\ \vspace{1.5mm}
\hspace{6.5mm}A_{4} \hspace{3mm} B_{4}\hspace{3mm} C_{4}\hspace{3mm} 
D_{4}\hspace{3mm} E_{4} \\  \vspace{1.5mm}
\hspace{24mm} \ddots    \\  \vspace{1.5mm}
\hspace{8mm} A_{n-2}\hspace{3mm} B_{n-2}\hspace{3mm} C_{n-2}\hspace{3mm} 
D_{n-2}\hspace{3mm}   E_{n-2}   \\  \vspace{1.5mm}
\hspace{18mm} A_{n-1}\hspace{3mm} B_{n-1}\hspace{3mm} C_{n-1}\hspace{3mm} 
D_{n-1}\hspace{3mm}  E_{n-1}   \\  \vspace{1.5mm}
\hspace{30mm} A_{n}\hspace{3mm} \ \ \ B_{n}\hspace{3mm} \ \ \  
C_{n}\hspace{3mm} \ \ \ D_{n}\hspace{3mm} \ \  E_{n}\hspace{3mm}   
\\ \vspace{1.5mm}
\hspace{49mm} -2\hspace{3mm}  \ \ \ -3  \hspace{3mm} \ \ \  
6 \hspace{3mm} \ \ \ \ -1 \\  \vspace{1.5mm}
\hspace{48mm} A_{n+2} \hspace{3mm} B_{n+2} \hspace{3mm}  C_{n+2}\hspace{3mm}  
D_{n+2} \hspace{3mm}  E_{n+2} \\ \vspace{1.5mm}
\hspace{58mm} A_{n+3}\hspace{3mm}   B_{n+3}\hspace{3mm}    C_{n+3}\hspace{3mm}   
D_{n+3}\hspace{3mm}   E_{n+3}\\ \vspace{1.5mm}
\hspace{68mm}A_{n+4}\hspace{3mm}     B_{n+4}\hspace{3mm}   C_{n+4}\hspace{3mm}   
D_{n+4}\hspace{3mm}   E_{n+4}  \\ \vspace{1.5mm}
\hspace{100mm}         \ddots  \\ \vspace{1.5mm}
\hspace{84mm} A_{2n-2} \hspace{3mm}   B_{2n-2} \hspace{3mm}   
C_{2n-2} \hspace{3mm}  
D_{2n-2} \hspace{3mm}   E_{2n-2}   \\ \vspace{1.5mm}
\hspace{95mm}  A_{2n-1} \hspace{3mm}   B_{2n-1} \hspace{3mm}   
C_{2n-1}\hspace{3mm}   
D_{2n-1} \hspace{3mm} E_{2n-1}\\ \vspace{1.5mm}
\hspace{5.5mm} E_{2n} \hspace{97mm}  A_{2n} \hspace{3mm} \ \ \  
B_{2n} \hspace{3mm} 
\ \ \ C_{2n} \hspace{3mm} \ \ \ D_{2n} \\ \vspace{1.5mm}
\hspace{6mm}  -2\hspace{106mm}  \ \ \ \ -1  \hspace{3mm} \ \ \ \ \ 6 
\hspace{3mm} \ \ \ \ -3 
\end{array}
\right),
}
\end{eqnarray}
\end{widetext}
where $A_i$, $B_i$, $C_i$, $D_i$ and $E_i$ are the coefficients of 
the discretized equations of Eq.~\ref{eq:eq1DFinal}. 
In the first row of \textbf{U}, the boundary condition $\Psi(0)=\Psi(D)$ 
(Eq.~\ref{conda}) is explicitly written as $\Psi_1=\Psi_{2n+1}$ while 
in the last row, the discretization of $\left. \frac{\partial}{\partial z} 
\Psi(z)\right|_{z=D} = 0$ (Eq.~\ref{condb}) is implemented as 
$-2\Psi_2-\Psi_{2n-1}+6\Psi_{2n}-3\Psi_{2n+1} = 0$. 
The periodic property of the solutions is taken into consideration in 
the last row ($\Psi_{2n+2}=\Psi_{2}$), in the second row (see $A_2$) and 
in the second last row (see $E_{2n}$). 

It is easy to verify that only the $C_i$ coefficients contain 
the magnetic field $B$ and these coefficients can be separated into 
two terms, where one of them is $B$-dependent\cite{Wang02b},
\begin{equation}
C_{i} = C^0_i + C^B_i \cdot B^2. 
\label{eq:cc}
\end{equation}
To obtain $B_{c2}$ with our method, we require that there be no prefactor 
appearing before the square of the magnetic field $B^2$. 
Hence, the factor $C^B_i$ in Eq.~\ref{eq:cc} should be divided. 
To cure the division singularity at $z=D/2$ 
(i.e., $C^B_{i}=0$ at $i=n+1$ in Eq.~\ref{eq:cc}), a solution property 
$\left. \frac{\partial}{\partial z} \Psi(z)\right|_{z=D/2} =0$ 
is assumed and implemented in the middle row of 
\textbf{U} as $-2\Psi_{n}-3\Psi_{n+1}+6\Psi_{n+2}-\Psi_{n+3} = 0$ 
(different from the approximation in the last row).

For Eq.~\ref{Ueq} to have non-trivial solutions, the determinant of 
$\textbf{U}$ should be zero,
\begin{equation}
\text{det}|\textbf{U}|=0. 
\end{equation}
By eliminating the constant elements in the first, middle and last rows of 
\textbf{U} and separating $C_i$ into the two terms with 
the negative prefactor $-C^B_i$ excluded (cf. Eq.~\ref{eq:cc}), 
det$|\textbf{U}|$ can be transformed into 
\begin{equation}
\text{det}|\textbf{P}-B^{2}\textbf{I}|=0, 
\end{equation}
where $\textbf{I}$ is a unitary matrix and $\textbf{P}$ is a matrix without 
the magnetic field (see below). Thus, the largest solution for $B$, 
namely $B_{c2}$,  can be easily available just by obtaining 
the largest positive eigenvalue of the eigen equation,
\begin{equation}
\textbf{P}\bbox{\Phi}=B^{2}\bm{\Phi}, 
\label{Peq}
\end{equation}
where \bm{$\Phi$} is the eigen function and \textbf{P}, with the dimension of 
$(2n-2)\times(2n-2)$, has the following structure, 
\begin{widetext}
\begin{eqnarray}
{\tiny
\left(
\begin{array}{l}
\vspace{3mm}
{C^0_{2}}^\prime-\frac{2}{3}B^\prime_2\hspace{4mm} D^\prime_{2}\hspace{3mm} 
E^\prime_{2} \hspace{86mm} -\frac{1}{3}B^\prime_{2}\hspace{15mm}  
A^\prime_{2}+2B^\prime_{2} \\ 
\vspace{3mm} 
B^\prime_{3}-\frac{2}{3}A^\prime_{3}\hspace{5mm} {C^0_{3}}^\prime\hspace{2mm} 
D^\prime_{3}\hspace{3mm} E^\prime_{3}  \hspace{80mm} -\frac{1}{3}A^\prime_{3} 
\hspace{18mm}  2A^\prime_{3} \\ 
\vspace{3mm}
\hspace{4mm}A^\prime_{4}\hspace{8mm} \ B^\prime_{4}\hspace{4mm} 
{C^0_{4}}^\prime\hspace{2mm} D^\prime_{4}\hspace{2mm} E^\prime_{4} \\  
\vspace{3mm}
\hspace{7mm} \ddots \hspace{13mm} \ddots \hspace{13mm} \ddots    \\  
\vspace{3mm}
\hspace{8mm} A^\prime_{n-2}\hspace{2mm}   B^\prime_{n-2} \hspace{2mm}  
{C^0_{n-2}}^\prime\hspace{2mm}   D^\prime_{n-2}\hspace{8mm}   E^\prime_{n-2} 
\\  \vspace{3mm}
\hspace{17mm} A^\prime_{n-1}\hspace{2mm}    B^\prime_{n-1}  \hspace{2mm}  
{C^0_{n-1}}^\prime\hspace{3mm}    D^\prime_{n-1}-\frac{2}{3}E^\prime_{n-1} 
\hspace{7mm} \  2E^\prime_{n-1} \hspace{10mm} \   -\frac{1}{3}E^\prime_{n-1} 
\\  \vspace{3mm}
\hspace{28mm} A^\prime_{n} \hspace{5mm}  B^\prime_{n} \hspace{7mm} \   
{C^0_{n}}^\prime-\frac{2}{3}D^\prime_{n} \hspace{8mm}  
E^\prime_{n}+2D^\prime_{n} \hspace{10mm} -\frac{1}{3}D^\prime_{n}   \\  
\vspace{3mm}
\hspace{45mm} A^\prime_{n+2}-\frac{2}{3}B^\prime_{n+2} \hspace{4mm}  
{C^0_{n+2}}^\prime+2B^\prime_{n+2} \hspace{3mm}  
D^\prime_{n+2}-\frac{1}{3}B^\prime_{n+2} \hspace{5mm}  E^\prime_{n+2} \\ 
\vspace{3mm}
\hspace{48mm} -\frac{2}{3}A^\prime_{n+3}\hspace{9mm} 
^\prime_{n+3}+2A^\prime_{n+3}\hspace{4mm}    
{C^0_{n+3}}^\prime-\frac{1}{3}A^\prime_{n+3}\hspace{5mm}   
D^\prime_{n+3}\hspace{3mm}   E^\prime_{n+3}\\ 
\vspace{3mm}
\hspace{72mm}A^\prime_{n+4}\hspace{14mm}   \  B^\prime_{n+4}\hspace{11mm}   
{C^0}^\prime_{n+4}\hspace{2mm}   D^\prime_{n+4}\hspace{2mm}   E^\prime_{n+4}  \\ 
\vspace{3mm}
\hspace{90mm} \ddots \hspace{13mm} \ddots \hspace{13mm} \ddots  \\ 
\vspace{3mm}
\hspace{76mm} A^\prime_{2n-2} \hspace{2mm}   B^\prime_{2n-2} \hspace{2mm}   
{C^0}^\prime_{2n-2} \hspace{2mm}  \ \ \ \ \ \ D^\prime_{2n-2} \hspace{15mm} \   
E^\prime_{2n-2}   \\ 
\vspace{3mm}
 -\frac{2}{3}E^\prime_{2n-1}\hspace{74mm}   A^\prime_{2n-1} \hspace{2mm}   
B^\prime_{2n-1} \hspace{3mm}   {C^0_{2n-1}}^\prime-\frac{1}{3}E^\prime_{2n-1} 
\hspace{3mm}  D^\prime_{2n-1}+2E^\prime_{2n-1} \\
\vspace{3mm}
E^\prime_{2n}-\frac{2}{3}D^\prime_{2n} \hspace{81mm}  \ \  A^\prime_{2n} 
\hspace{4mm} \ \ \ \   B^\prime_{2n}-\frac{1}{3}D^\prime_{2n} \hspace{3mm}  
\ \ \ \ \ \ \  {C^0_{2n}}^\prime+2D^\prime_{2n}  
\end{array}
\right),
\vspace{20mm}
}
\end{eqnarray}
\end{widetext}
where $A_i^\prime=-A_i/C^B_i$, $B_i^\prime=-B_i/C^B_i$, 
${C^0_i}^\prime=-C^0_i/C^B_i$, $D_i^\prime=-D_i/C^B_i$ and 
$E_i^\prime=-E_i/C^B_i$. Here, $i\not=n+1$ and thus   
the division singularity is avoided. Having determined 
$B_{c2}$ from Eq.~\ref{Peq},  one can obtain 
the corresponding order parameter by substituting $B_{c2}$ back 
into Eq.~\ref{Ueq}. 

\subsection{Procedure II}

In the above procedure, the upper critical field is obtained by treating 
the square of the magnetic field $B^2$ as an eigenvalue (Eq.~\ref{Peq}). 
Following the same idea, one may obtain a straight forward procedure from 
Eq.~\ref{eq:eq1DFinal}: 
\begin{equation}
\textbf{Q}\bm{\Psi}=B^{2}\bm{\Psi}, 
\label{Qeq}
\end{equation}
where $\bm{\Psi} = (\Psi_1, \Psi_2, ..., \Psi_{2n})^\prime$, 
which has the same meaning as that in Eq.~\ref{Ueq}. However, to avoid 
the singularity at $z=D/2$, the dimension of the current \bm{$\Psi$}, $2n$, 
is set different from the previous one, $2n+1$. 
The matrix \textbf{Q} has the following structure\cite{Wang02b}, 
\begin{widetext}
\begin{eqnarray}
{\footnotesize 
\left(
\begin{array}{l} \vspace{1.5mm}
C^0_1\hspace{6mm}D_1\hspace{6mm}E_1  \hspace{52mm} A_1 \hspace{11mm} B_1 
\\ \vspace{1.5mm}
B_{2}\hspace{6mm} C^0_{2}\hspace{6mm} D_{2}\hspace{6mm} E_{2} \hspace{57mm} A_{2}  
\\ \vspace{1.5mm}
A_{3}\hspace{6mm} B_{3}\hspace{6mm} C^0_{3}\hspace{5.5mm} D_{3}\hspace{6mm} E_{3}  
\\  \vspace{1.5mm}
\hspace{10mm}A_{4} \hspace{6mm} B_{4}\hspace{6mm} C^0_{4}\hspace{6mm} 
D_{4}\hspace{6mm} 
E_{4} \\  \vspace{1.5mm}
\\ \vspace{1.5mm}
\hspace{45mm} \ddots    \\  \vspace{1.5mm}
\\ \vspace{1.5mm}
\hspace{29mm} A_{2n-3} \hspace{6mm}   B_{2n-3} \hspace{6mm}   
C^0_{2n-3} \hspace{5.5mm}  D_{2n-3} \hspace{6mm}   E_{2n-3}   \\ \vspace{1.5mm}
\hspace{44mm}  A_{2n-2} \hspace{6mm}   B_{2n-2} \hspace{6mm}   
C^0_{2n-2}\hspace{6mm}   
D_{2n-2} \hspace{6mm} E_{2n-2}\\ \vspace{1.5mm}
\hspace{7mm} E_{2n-1} \hspace{43mm}  A_{2n-1} \hspace{6mm}   
B_{2n-1} \hspace{6mm} 
C^0_{2n-1} \hspace{6mm}  D_{2n-1} \\ \vspace{1.5mm}
\hspace{9mm}  D_{2n}\hspace{4.5mm} E_{2n} \hspace{48mm}  \ \ \  
A_{2n}  \hspace{6.5mm}  
\ \ \  B_{2n} \hspace{6mm} \ \ \   C^0_{2n}
\end{array}
\right).
}
\end{eqnarray}
\end{widetext}
The dimension of \textbf{Q} is $2n\times2n$, larger than that of 
\textbf{P} but smaller than that of \textbf{U}. 

Eq.~\ref{Qeq} implies that one may directly determine 
the upper critical field from the linear GL equation. 
The present procedure is essentially consistent with 
the usual method for obtaining $B_{c2}$ through the smallest eigenvalue of 
the linear GL equation since there exists an inverse relation between 
the magnetic field eigenvalues and the eigenvalues 
(or the indices of the Landau levels) of the linear GL equation 
(for example, see Ref.~\onlinecite{Duzer99}). 
Our procedures (particularly Procedure II) indicate that 
one may directly determine the upper critical field without 
considering the smallest eigenvalue of an appropriate eigen equation. 
In fact, in one of the earliest papers\cite{Gorkov60}, 
Gor'kov simply used a variational method to obtain $B_{c2}$. 

\section{Results and Discussion}

Bi2212 has been chosen as our modeling prototype since it possesses 
a large anisotropy\cite{Farrell89}, and is thus suitable for studies of 
properties related to the layered structure. The values of the parameters 
for Bi2212 are taken from Ref.~\onlinecite{Wang01b}. 
The calculated $B_{c2}$ at zero temperature for various $n$ are 
shown in Table~\ref{tab:Bc2Bi2212}. It can be seen that the two sets of 
data calculated from Eqs.~\ref{Peq} and \ref{Qeq} converge finally and 
the convergent values are in good agreement with each other. 

\begin{table}[b]
\caption{
Values of $B_{c2}$ (SI unit) for various $n$, treated as eigenvalue problems 
from different procedures. The error is $E_n=|B_{c2}(n)-B_{c2}(n=1200)|$, 
where $B_{c2}(n=1200)=3338.680126$ Tesla. 
}
\begin{ruledtabular}
\begin{tabular}{lccr}
eigen equation  			& n  	  & $B_{c2}$      &  $E_n$    \\
\hline
       				& 50 	  & 3265.749101   & 72.93   	\\
		     			& 100   & 3328.453273   & 10.23   	\\
Eq.~\ref{Peq} (Procedure I)  	& 200   & 3337.366302 	& 1.314    	\\
					& 400	  & 3338.513958   & 0.1662    \\
					& 800   & 3338.658438   & 0.02169   \\
\hline
					& 50	  & 3339.154981	& 0.4749    \\
					& 100	  & 3338.709744	& 0.02962   \\
Eq.~\ref{Qeq} (Procedure II)  & 200	  & 3338.681966   & 0.001840  \\
					& 400	  & 3338.680240	& 1.135E-4  \\
					& 800	  & 3338.680132	& 6.020E-6  \\
\end{tabular} \label{tab:Bc2Bi2212}
\end{ruledtabular}
\end{table}

\begin{figure}[t]
\includegraphics[bb=101 300 461 672,scale=0.65,clip]{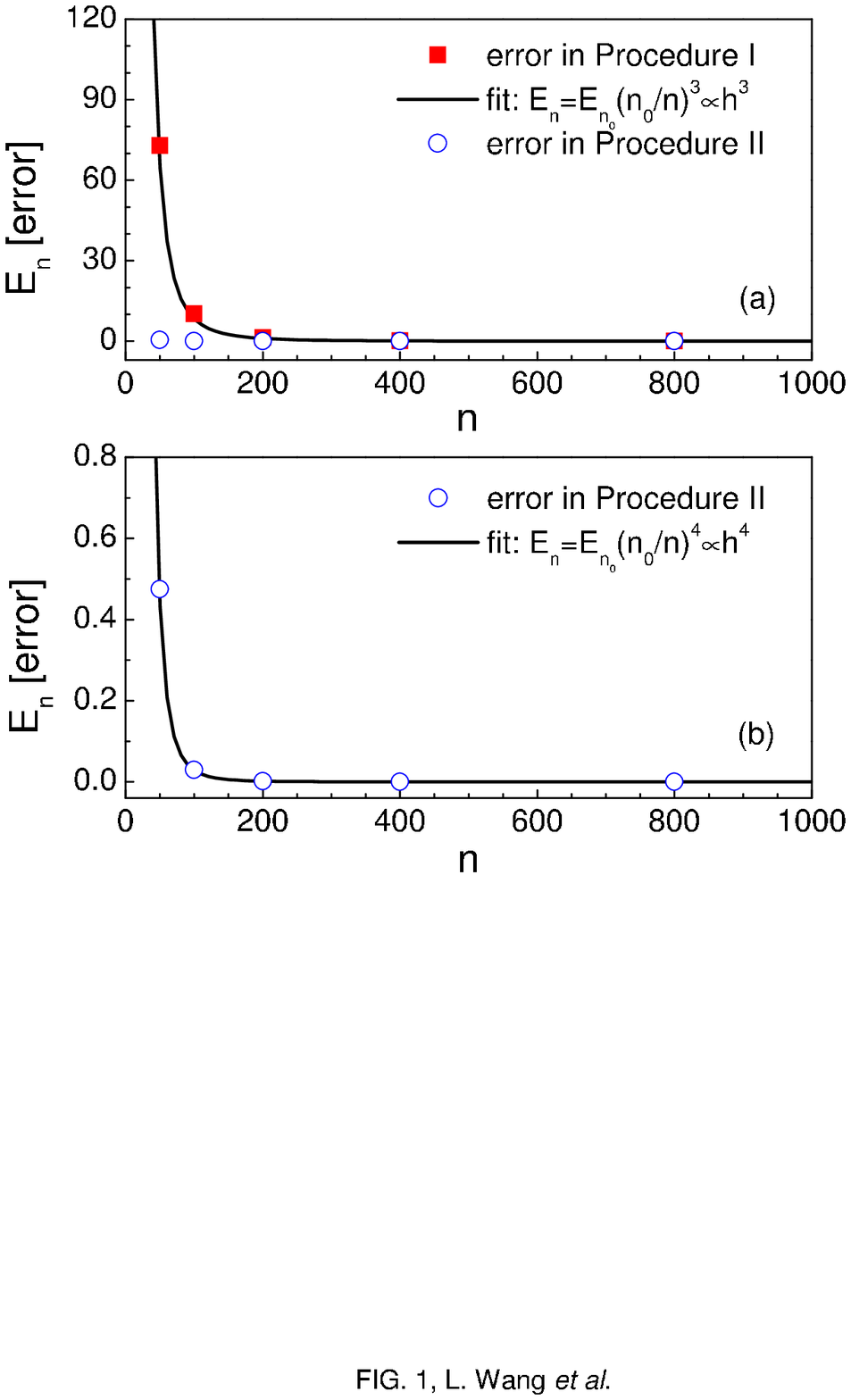}
\caption{{\footnotesize
Errors for Procedure I and Procedure II, which are of order 3 and 4, 
respectively. 
}}
\label{fig:error}
\end{figure}

A glance at the data for $B_{c2}$ in Table~\ref{tab:Bc2Bi2212} shows that 
Procedure II converges more rapidly than Procedure I. 
Hence, using $B_{c2}(n=1200)$ from Procedure II for 
the convergence analysis, the error is defined as 
\begin{equation}
E_n= |B_{c2}(n)-B_{c2}(n=1200)|.
\end{equation}
This equation is applicable to both procedures and the corresponding errors  
are listed in Table~\ref{tab:Bc2Bi2212} and plotted in Fig.~\ref{fig:error}.  
It is found that the errors of the two procedures can be 
respectively fitted as 
\begin{equation}
E_n = \left\{ 
\begin{array}{ll}
E_{n_0}(n_0/n)^3 \propto h^3 & \ \ \ \text{for Procedure I}, \\ 
E_{n_0}(n_0/n)^4 \propto h^4 & \ \ \ \text{for Procedure II}. 
\end{array} 
\right.
\label{eq:eqerror}
\end{equation}
Here $n_0$ is taken to be 50 and $E_{n_0}$ is the corresponding error. 
The grid spacing $h \sim D/2n$, when $n$ is large. From these fits, 
we know that the error of Procedure I is of order 3 while that of 
Procedure II is 4, which are consistent with the error orders of 
the approximations used in the matrices \textbf{U} and \textbf{Q}, 
respectively. 

Note that $B_{c2}$ at 0 K in the tables may be comparable to 
those extrapolated (using the WHH theory\cite{Werthamer66}) from 
some experiments (for example, see Ref.~\onlinecite{Palstra88}). 
By choosing appropriate values of the model parameters, 
the extrapolated result of 2640 Tesla for Bi2212\cite{Palstra88} 
can be obtained exactly.

\begin{figure}[t]
\includegraphics[bb=103 382 450 673,scale=0.6,clip]{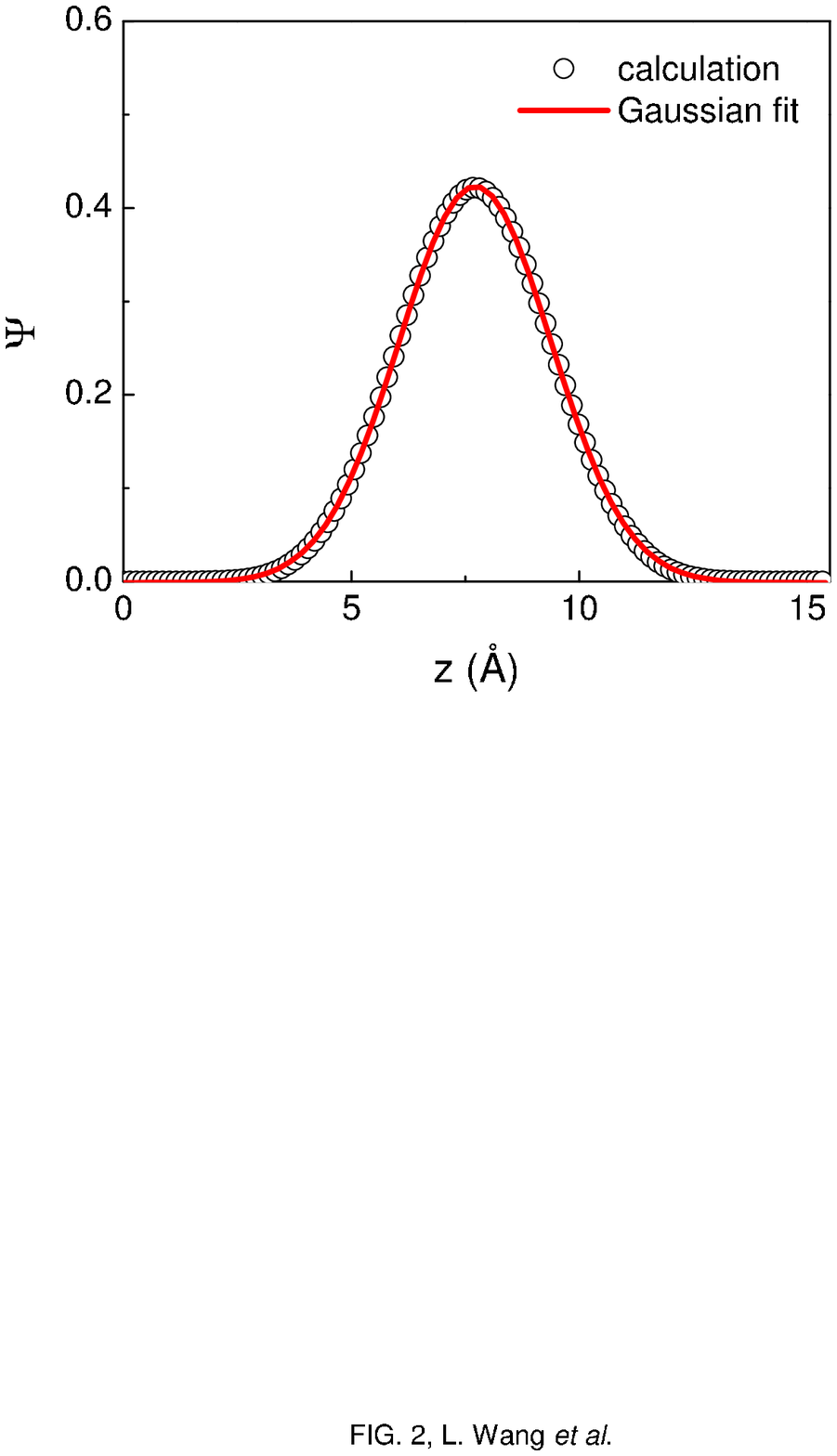}
\caption{{\footnotesize
Spatial distribution of the order parameter at $T/T_c = 0.9$, 
which can be approximated by the well-known Gaussian function.
}}
\label{fig:Gauss}
\end{figure}

We find that the order parameters obtained from Eqs.~\ref{Ueq} and \ref{Qeq} 
are also consistent with each other. In the following calculations, 
Procedure II is employed. In Fig.~\ref{fig:Gauss}, we plot 
the spatial distribution of the calculated order parameter at 
$T/T_c=0.9$. It is found that for Bi2212 in a large temperature range, 
the asymptotic behavior of the order parameter can be expressed as 
\begin{equation}
\Psi(z) \sim C\exp\left[-\frac{(z-D/2)^2}{2\xi^2}\right],
\label{eq:Gauss}
\end{equation}
which is exactly the ground state of the linearized GL equation at 
$B_{c2}$\cite{Abrikosov57,Tinkham96}. The Gaussian profile of 
the order parameter is also similar to the shape of 
the pair amplitude in bulk superconductors. 
In fact, it is natural that the order parameter of 
the largest eigenvalue of $B^2$ should be just that of 
the lowest eigenvalue for the linear equation, 
due to the inverse relation between $B$ and the eigenvalues of 
the linear equation\cite{Tinkham96,Duzer99}. 

When near $T_c$ or in a less anisotropic superconductor such as YBCO, 
we found that an offset of the Gaussian order parameter (Eq.~\ref{eq:Gauss}) 
can not be neglected. Hence, one may infer that the mean field effect or 
the interlayer coupling\cite{Wang01a} of the order parameter 
has less influence on a highly anisotropic superconductor 
at low temperatures (2-D feature) than on a superconductor at 
high temperatures or with a small anisotropy\cite{Wang01b}.  

\begin{figure}[b]
\includegraphics[bb=103 382 450 673,scale=0.6,clip]{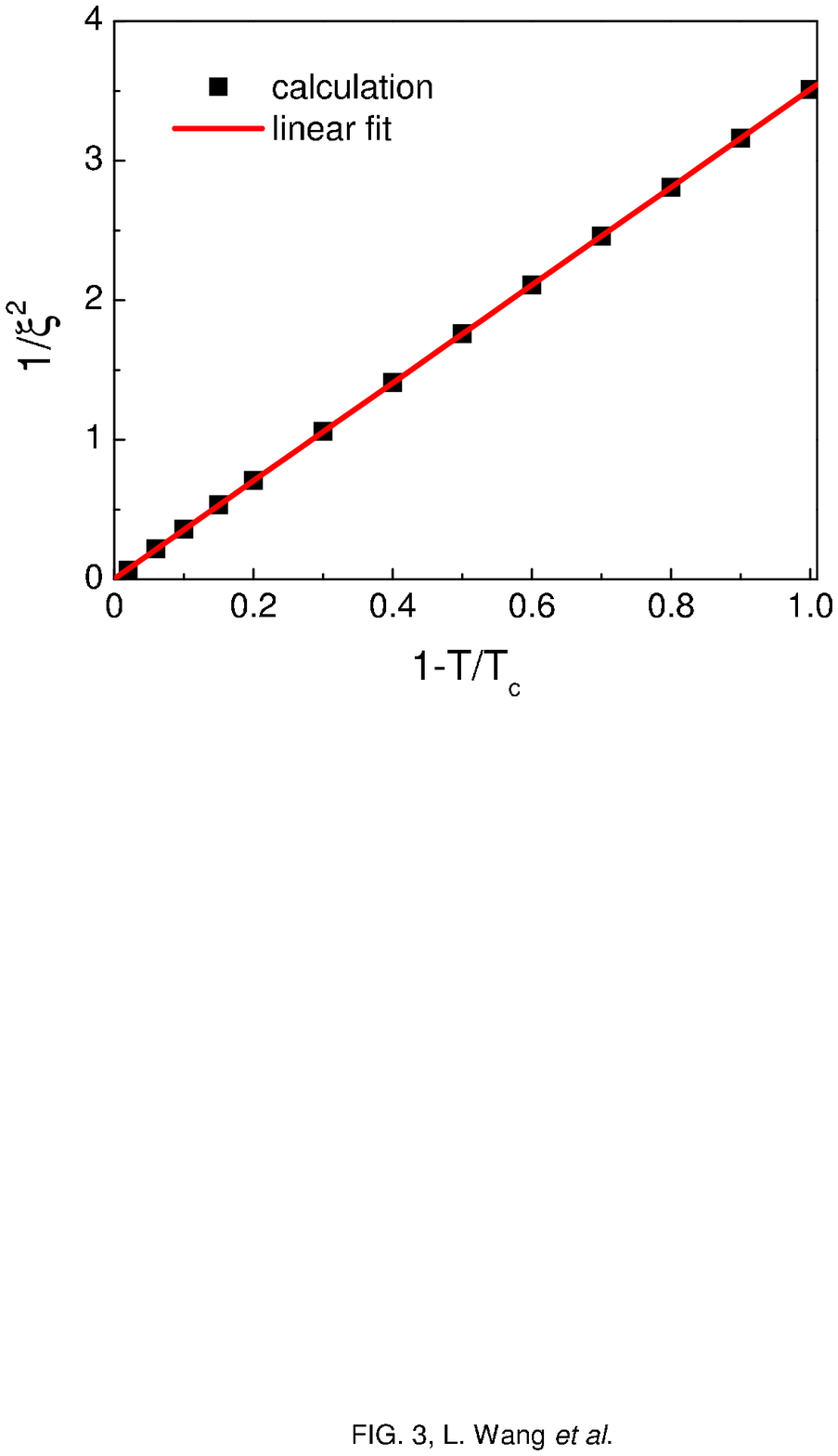}
\caption{{\footnotesize
Linear relation between $1/\xi^2$ and $1-T/T_c$, where $\xi$ is 
the parameter characterizing the width of the Gaussian order parameter.
}}
\label{fig:linear}
\end{figure}

\begin{figure}[t]
\includegraphics[bb=103 382 450 673,scale=0.6,clip]{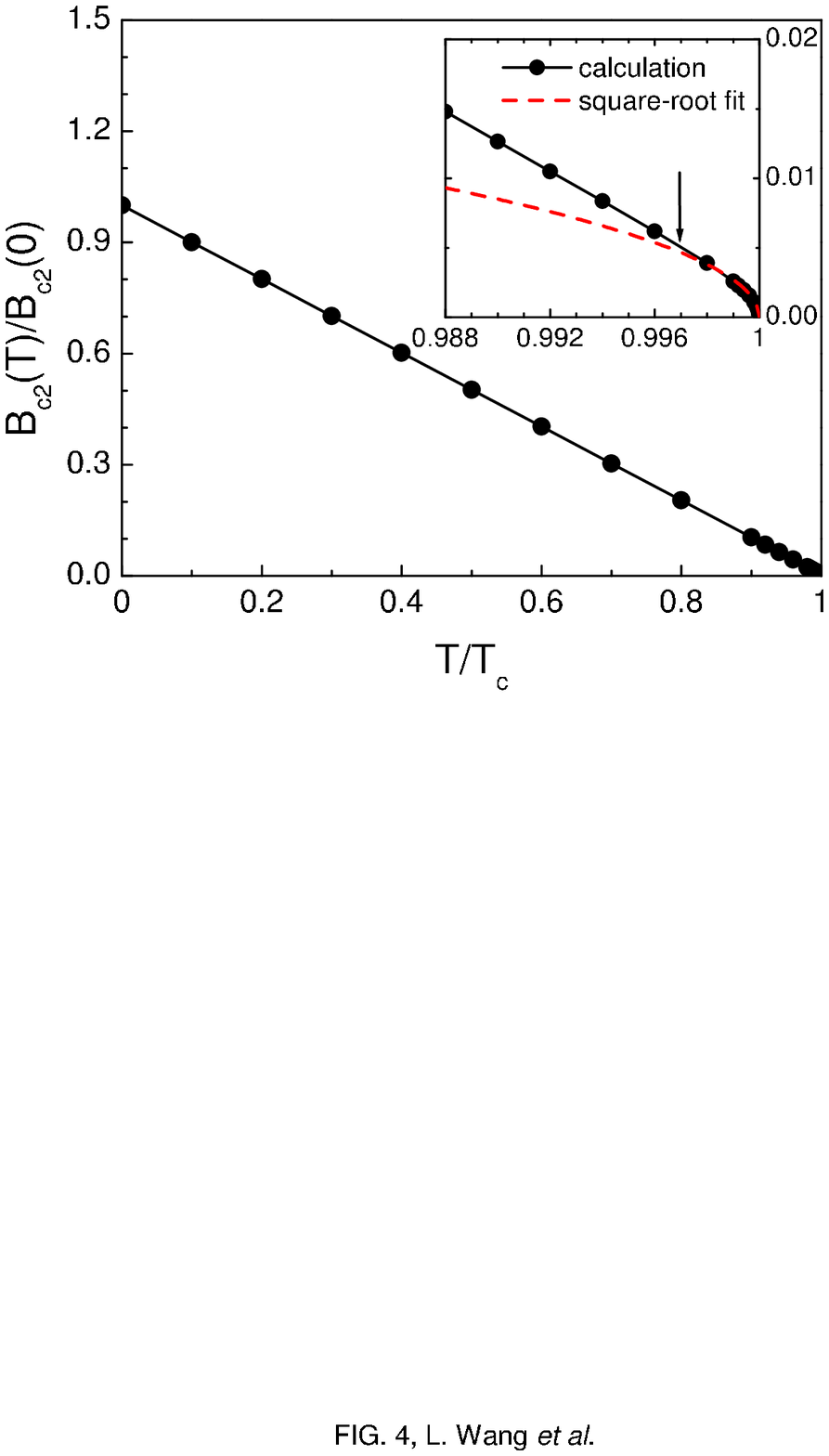}
\caption{{\footnotesize
The $B_{c2}$-$T$ behavior is linear in a large temperature range but 
square-root-like near $T_c$ (inset). The solid circles ($\bullet$)  
stand for the calculated data and the full lines are guidances to 
the eye. The dashed line represents a square-root fit. 
The reduced temperature of the transition point from square-root to 
linearity is about 0.997 (indicated by the arrow).
}}
\label{fig:Bc2}
\end{figure}

The finding here that a Gaussian-like exponential component 
contributes to the order parameter may adequately signify that 
the eigen function associated with the maximum magnetic field eigenvalue of 
the linear GL equation is equivalent, to a certain extent, to 
the eigen function related to the lowest eigenvalue of 
the linear GL equation. The latter eigen function may have 
the form $\exp\left[\frac{B_{c2}\pi}{\Phi_0}(z-D/2)^2\right]$ 
(cf. Eq.~4.73 in Ref.~\onlinecite{Tinkham96}). Comparing this function to 
the Gaussian-like exponential contribution (Eq.~\ref{eq:Gauss}), 
one has $B_{c2} \propto 1/\xi^2$. For Bi2212, we find that 
$1/\xi^2 \propto 1-T/T_c$ is satisfied for a large temperature range 
(Fig.~\ref{fig:linear}), hence, $B_{c2} \propto 1-T/T_c$ 
(see Fig.~\ref{fig:Bc2}). Similarly, we also have found that 
the linear $B_{c2}$-$T$ relation is satisfied in YBCO. 
Such a linear feature is consistent with 
the general $B_{c2}$-$T$ picture in layered cuprates near 
$T = 0$ K\cite{Kresin99}, and is in agreement with 
the anisotropic GL theory and some other 
theories\cite{Schneider93,Joynt90,Feinberg94,Theodorakis89}. 
Experimentally, the linear behavior away from $T_c$ was observed in 
YBCO\cite{Welp89,Moodera88}. 

A closer inspection of Fig.~\ref{fig:Bc2} reveals that near $T_c$, 
the $B_{c2}$-$T$ plot exhibits a square-root-like behavior which is 
in agreement with Feinberg's theory\cite{Feinberg94}. 
This behavior is also found in twined superconductors\cite{Fang88}, 
multilayers\cite{Radovic88}, and thin films\cite{Tinkham96,Averin83}. 
In one of the earliest papers, Ginzburg and Landau\cite{Ginzburg50} 
correctly predicted that $B_{c2}$ for thin films varied as 
$\sqrt{T_c-T}$ when $T \rightarrow T_c$, as a consequence of 
the boundary conditions. This was subsequently confirmed experimentally 
by Khukhareva\cite{Khukhareva62}. 

In our calculations, we found that near $T_c$, the height of 
the Gauss component $C$ in Eq.~\ref{eq:Gauss} decreases with temperature 
and hence, the order parameter is nearly constant 
(3-D feature)\cite{Wang02a,Wang02b}. Qualitatively speaking, 
the nearly constant order parameter near $T_c$ for 
the layered superconductors is equivalent to the case of 
thin films so that one can observe the square-root behavior 
near $T_c$. In fact, the boundary condition of Eq.~\ref{condb} 
does apply to thin films. More precisely, with a constant solution to 
the order parameter, one can immediately infer from Eq.~\ref{eq:eq1DFinal} 
that there exists a square-root relationship between $B_{c2}$ and $1-T/T_c$. 
It is worth noting that Dediu {\it et al}. also obtained 
a square-root $B_{c2}-T$ relation near $T_c$ by solving 
the usual GL equations and the feature of the order parameter 
near $T_c$ was used to interpret their results.

It is interesting to note that in the $B_{c2}$-$T$ plot, 
the transition from square-root to linearity may be considered as 
a 3-D to 2-D crossover. The transition temperature obtained 
here (indicated by the arrow in Fig.~\ref{fig:Bc2}) is 
reasonably consistent with the estimates in 
Refs.~\onlinecite{Tinkham96} and \onlinecite{Sokolov91}.

\section{Conclusion}

The conventional work to determine upper critical fields 
should resort to obtaining the smallest eigenvalue of an appropriate 
eigen equation. However, within a continuous Ginzburg-Landau model, 
we have demonstrated through two procedures that one can obtain 
the upper critical field by treating the square of the magnetic field 
as eigenvalue problems, from which $B_{c2}$ can be directly deduced. 

The calculated $B_{c2}$ from the two procedures are consistent 
with each other and in reasonably good agreement with 
existing theories and experiments. The profile of 
the order parameter obtained at $B_{c2}$ is Gaussian-like, 
further indicating the plausibility of the procedures proposed. 

The convergences of the proposed procedures were also investigated 
to identify the efficiency of the procedures. It was found that 
the more direct method, Procedure II, converges faster than 
Procedure I. The corresponding convergent error orders of the two 
procedures are 4 and 3, respectively. These orders are consistent with 
the orders of the approximations used in our calculations.  

Note that certain physical phenomena such as fluctuations and 
spin-orbit scattering may have some influences on 
the upper critical field. The present procedures proposed may   
serve as a starting point to further study more properties of 
upper critical fields.


\begin{references}

\bibitem{Ginzburg50} V. L. Ginzburg and L. D. Landau, 
Zh. Eksp. Teor. Fiz. {\bf 20}, 1064 (1950). 

\bibitem{Abrikosov57} A. A. Abrikosov, Zh. Eksp. Teor. Fiz. 
{\bf 32}, 1442 (1957) [Sov. Phys.-JETP {\bf 5}, 1174 (1957)].

\bibitem{Gorkov59} L. P. Gor'kov, Zh. Eksp. Teor. Fiz. {\bf 36}, 1918 (1959) 
[Sov. Phys.-JETP {\bf 9}, 1364 (1959)]; 
{\bf 37}, 1407 (1959) [{\bf 10}, 998 (1960)].

\bibitem{Cheng99} S. L. Cheng and H. A. Fertig, Phys. 
Rev. B {\bf 60}, 13107 (1999).

\bibitem{Tinkham96} M. Tinkham, {\it Introduction to Superconductivity} 
(Mcgraw-Hill, Inc., 1996).

\bibitem{Duzer99} T. V. Duzer and C. W. Turner, {\it Principles of 
Superconductive Devices and Circuits} (Prentice-Hall, Inc. 1999).

\bibitem{Jin89} B. Y. Jin and J. B. Ketterson, 
Adv. Phys. {\bf 38}, 189 (1989). 

\bibitem{Bulaevskii76} L. N. Bulaevskii, Usp. Fiz. Nauk {\bf 116}, 449 (1975) 
[Sov. Phys.-Usp. {\bf 18}, 514 (1976)]. 

\bibitem{Schneider93} T. Schneider and A. Schmidt,
Phys. Rev. B {\bf 47}, 5915 (1993).  

\bibitem{Klemm75} R. A. Klemm, A. Luther, and M. R. Beasley, 
Phys. Rev. B {\bf 12}, 877 (1975).

\bibitem{Koyama92} T. Koyama, N. Takezawa, Y. Naruse, and M. Tachiki, 
Physica C {\bf 194}, 20 (1992).

\bibitem{Joynt90} R. Joynt, Phys. Rev. B {\bf 41}, 4271 (1990). 

\bibitem{Berlinsky95} A. J. Berlinsky, A. L. Fetter, M. Franz, C. Kallin, 
and P. I. Soininen, Phys. Rev. Lett. {\bf 75}, 2200 (1995).

\bibitem{Franz96} M. Franz, C. Kallin, P. I. Soininen, A. J. Berlinsky, 
and A. L. Fetter, Phys. Rev. B {\bf 53}, 5795 (1996). 

\bibitem{Chang98} D. Chang, C. Y. Mou, B. Rosenstein, and C. L. Wu, 
Phys. Rev. B {\bf 57}, 7955 (1998). 

\bibitem{Alexandrov96} A. S. Alexandrov, Phys. Rev. B {\bf 48}, 10571 (1993);
A. S. Alexandrov, V. N. Zavaritsky, W. Y. Liang, and P. L. Nevsky, 
Phys. Rev. Lett. {\bf 76}, 983 (1996).

\bibitem{Gennes64} P. G. de Gennes, Rev. Mod. Phys. {\bf 36}, 225 (1964).

\bibitem{Werthamer63} N. R. Werthamer, Phys. Rev. {\bf 132}, 2440 (1963).

\bibitem{Maki64} K. Maki, Physics {\bf 1}, 21 (1964); 
Phys. Rev. {\bf 148}, 362 (1966).

\bibitem{Werthamer66} E. Helfand and N. R. Werthamer, 
Phys. Rev. {\bf 147}, 288 (1966); 
N. R. Werthamer, E. Helfand, and P. C. Hohenberg, 
Phys. Rev. {\bf 147}, 295 (1966).

\bibitem{Scharnberg80} K. Scharnberg and R. A. Klemm, 
Phys. Rev. B {\bf 22}, 5233 (1980). 

\bibitem{Lukyanchuk87} I. A. Luk'yanchuk and V. P. Mineev, 
Zh. Eksp. Teor. Fiz. {\bf 93}, 2030 (1987) 
[Sov. Phys.-JETP {\bf 66}, 1168 (1987)]. 

\bibitem{Rieck89} C. T. Rieck, Th. W\"{o}lkhausen, D. Fay, and L. Tewordt,  
Phys. Rev. B {\bf 39}, 278 (1989).

\bibitem{Maki97} K. Maki and M. T. Beal-Monod, 
Phys. Rev. B {\bf 55}, 11730 (1997).

\bibitem{Ovchinnikov96} Yu. N. Ovchinnikov and V. Z. Kresin, 
Phys. Rev. B {\bf 54}, 1251 (1996).

\bibitem{Lebed98} A. G. Lebed and K. Yamaji, 
Phys. Rev. Lett. {\bf 80}, 2697 (1998); 
A. G. Lebed, J. Supercond. {\bf 12}, 453 (1999).

\bibitem{Takahashi86} S. Takahashi and M. Tachiki, 
Phys. Rev. B {\bf 33}, 4620 (1986).

\bibitem{Yuan91} B. J. Yuan and J. P. Whitehead, 
Phys. Rev. B {\bf 44}, 6943 (1991). 

\bibitem{Yuan94} B. J. Yuan and J. P. Whitehead, 
Physica C {\bf 231}, 395 (1994). 

\bibitem{Smith00} R. A. Smith, B. S. Handy, and V. Ambegaokar, 
Phys. Rev. B {\bf 61}, 6352 (2000). 

\bibitem{Kuwasawa96} Y. Kuwasawa, T. Nojima, S. Hwang, 
B. J. Yuan, and J. P. Whitehead, Physica B {\bf 222}, 92 (1996). 

\bibitem{Huang89} X. Z. Huang and K. Maki, 
Phys. Rev. B {\bf 39}, 6459 (1989). 

\bibitem{exception} In Ref.~\onlinecite{Alexandrov96}, the first nontrivial 
extended solution to the Ginzburg-Pitaevskii equation appears at $B=B_{c2}$ 
when the chemical potential coincides with the energy of the lowest 
extended state (i.e., the lowest delocalized energy). The chemical potential 
may be simply defined as the minimal energy that is needed to add to 
a statistical system. Hence, when coinciding with 
the lowest delocalized energy, the chemical potential may be 
considered as ``the lowest eigenvalue" of the Ginzburg-Pitaevskii equation.

\bibitem{reenter} Note, however, that in Ref.~\onlinecite{Lebed98}, 
the linear Gor'kov gap equation may have different solutions, 
which may lead to the problem of the reentrant superconductivity.

\bibitem{Gorkov76} L. P. Gor'kov and N. B. Kopnin, 
Usp. Fiz. Nauk {\bf 116}, 413 (1975) 
[Sov. Phys.-Usp. {\bf 18}, 496 (1976)].

\bibitem{Wang01a} L. Wang, H. S. Lim, and C. K. Ong, 
Supercond. Sci. Technol. {\bf 14}, 252 (2001). 

\bibitem{Wang01b} L. Wang, H. S. Lim, and C. K. Ong, 
Supercond. Sci. Technol. {\bf 14}, 754 (2001). 

\bibitem{Wang02a} L. Wang, H. S. Lim, and C. K. Ong, 
Physica C {\bf 383}, 247 (2002).

\bibitem{Kleiner97} R. Kleiner and P. M\"{u}ller, 
Physica C {\bf 293}, 156 (1997).

\bibitem{Wang02b} L. Wang, Ph.D. thesis, 
National University of Singapore, 2002.

\bibitem{Gorkov60} L. P. Gor'kov, Zh. Eksp. Teor. Fiz. {\bf 37}, 833 (1959) 
[Sov. Phys.-JETP {\bf 10}, 593 (1960)].

\bibitem{Farrell89} D. E. Farrell, S. Bonham, J. Foster, Y. C. Chang, 
P. Z. Jiang, K. G. Vandervoort, D. J. Lam, and V. G. Kogan, 
Phys. Rev. Lett. {\bf 63}, 782 (1989).

\bibitem{Palstra88} T. T. M. Palstra, B. Batlogg, L. F. Schneemeyer, 
R. B. van Dover, and J. V. Waszczak, Phys. Rev. B {\bf 38}, 5102 (1988).

\bibitem{Kresin99} Vladimir Z. Kresin, Yu. N. Ovchinnikov, and Stuart A. Wolf,
J. Supercond. {\bf 12}, 493 (1999).    

\bibitem{Feinberg94} D. Feinberg, S. Theodorakis, and A. M. Ettouhami,
Phys. Rev. B {\bf 49}, 6285 (1994).      

\bibitem{Theodorakis89} S. Theodorakis and Z. Te\v{s}anovi\'{c}, 
Phys. Rev. B {\bf 40}, 6659 (1989).

\bibitem{Fang88} M. M. Fang, V. G. Kogan, D. K. Finnemore, J. R. Clem, 
L. S. Chumbley, and D. E. Farrell, Phys. Rev. B {\bf 37}, 2334 (1988). 

\bibitem{Radovic88} Z. Radovi\'{c}, L. Dobrosavljevi\'{c}-Gruji\'{c}, 
A. I. Buzdin, and J. R. Clem, Phys. Rev. B {\bf 38}, 2388 (1988). 

\bibitem{Averin83} V. V. Averin, A. I. Buzdin, and L. N. Bulaevskii, 
Zh. Eksp. Teor. Fiz. {\bf 84}, 737 (1983) 
[Sov. Phys.-JETP {\bf 57}, 426 (1983)].

\bibitem{Khukhareva62} I. S. Khukhareva, Zh. Eksp. Teor. Fiz. {\bf 41}, 728 (1961) 
[Sov. Phys.-JETP {\bf 14}, 526 (1962)].

\bibitem{Welp89} U. Welp, W. K. Kwok, G. W. Crabtree, K. G. Vandervoort, 
and J. Z. Liu, Phys. Rev. Lett. {\bf 62}, 1908 (1989). 

\bibitem{Moodera88} J. S. Moodera, R. Meservey, J. E. Tkaczyk, C. X. Hao, 
G. A. Gibson, and P. M. Tedrow, Phys. Rev. B {\bf 37}, 619 (1988).

\bibitem{Sokolov91} A. I. Sokolov, Physica C {\bf 174}, 208 (1991).

\end{references}
\end{document}